%% file: icann09.tex
\documentclass[runningheads,a4paper]{llncs}

\usepackage{amssymb}
\usepackage[dvips]{graphicx}

\newcommand{\keywords}[1]{\par\addvspace\baselineskip
\noindent\keywordname\enspace\ignorespaces#1}

\begin{document}
\input{CommandDefs.tex}
\mainmatter  

\title{Topographic Mapping of Astronomical Light Curves via a Physically Inspired Probabilistic Model}

\titlerunning{Topographic Mapping of Astronomical Light Curves}

%
%
\author{Nikolaos Gianniotis\inst{1} \and Peter Ti\v{n}o\inst{2} \and Steve Spreckley\inst{3} \and Somak Raychaudhury\inst{3}}
\authorrunning{Topographic Mapping of Astronomical Light Curves}

\institute{Heidelberg Collaboratory for Image Processing,\\
University of Heidelberg
D-69115 Heidelberg, Germany,
\and
School of Computer Science\\
The University of Birmingham\\
Edgbaston
B15 2TT
United Kingdom,
\and
School of Physics and Astronomy\\
The University of Birmingham\\
Edgbaston
B15 2TT
United Kingdom
}

%
%

\toctitle{Lecture Notes in Computer Science}
\tocauthor{Authors' Instructions}
\maketitle

\begin{abstract}
We present a probabilistic generative approach for constructing topographic maps 
of light curves from eclipsing binary stars.
The model defines a low-dimensional manifold of local noise models induced by a smooth non-linear
mapping from a low-dimensional latent space into the space of probabilistic models of the observed light curves. 
The local noise models are physical models that describe how such light curves are generated.
Due to the principled probabilistic nature of the model, a cost function arises
naturally and the model parameters are fitted via MAP estimation using the Expectation-Maximisation algorithm.
Once the model has been trained, each light curve may be projected
to the latent space as the the mean posterior probability over the local noise models.
We demonstrate our approach on a dataset of artificially generated light curves
and on a dataset comprised of light curves from real observations. 

\keywords{Topographic mapping, eclipsing binary stars}
\end{abstract}

\section{Introduction}

The Generative Topographic Map algorithm (GTM) \cite{Bishop1998}
has been  introduced as a probabilistic analog to SOM \cite{Kohonen1990}, seeking to address certain of 
its limitations such as the absence of a cost function. 
The GTM  formulates a mixture of spherical Gaussians densities constrained
on a smooth image of a low-dimensional latent space. Each point in the latent space 
is mapped via a smooth non-linear mapping to its image in the high-dimensional data space. This image plays the role of the mean of a local spherical Gaussian noise model that is responsible for modelling the density of data points in its vicinity.
The GTM can be readily extended to structured data by adopting
alternative formulations of noise models in the place of Gaussian densities. Such
extensions have been proposed in \cite{Tino2004} for the visualisation of symbolic 
sequences and in \cite{Gianniotis2007} for the visualisation of tree-structured data.

Here we present a further extension of the GTM to a novel data type, namely light curves that originate from eclipsing binary systems.
Binary stars are gravitationally bound pairs of stars that orbit a common
centre of mass. Astronomical observations suggest that almost half of the stars are
binary ones. 
Thus, studying such systems procures knowledge for a significant proportion of stars.
Binary stars are important to astrophysics 
because they allow calculation of fundamental quantities such as masses and radii,
and are important for the verification of theoretical models for stellar formation and evolution. 
A particular subclass of binary stars are eclipsing binary stars.
The luminosity of such stars varies over time and forms a graph called light curve. 
Light curves are important because they provide information on the characteristics of
stars and help in the identification of their type.

\section{Physical Model For Eclipsing Binaries}
\label{sec:physical_model}

\begin{figure*}[!t]
\centering
\includegraphics[width=6cm]{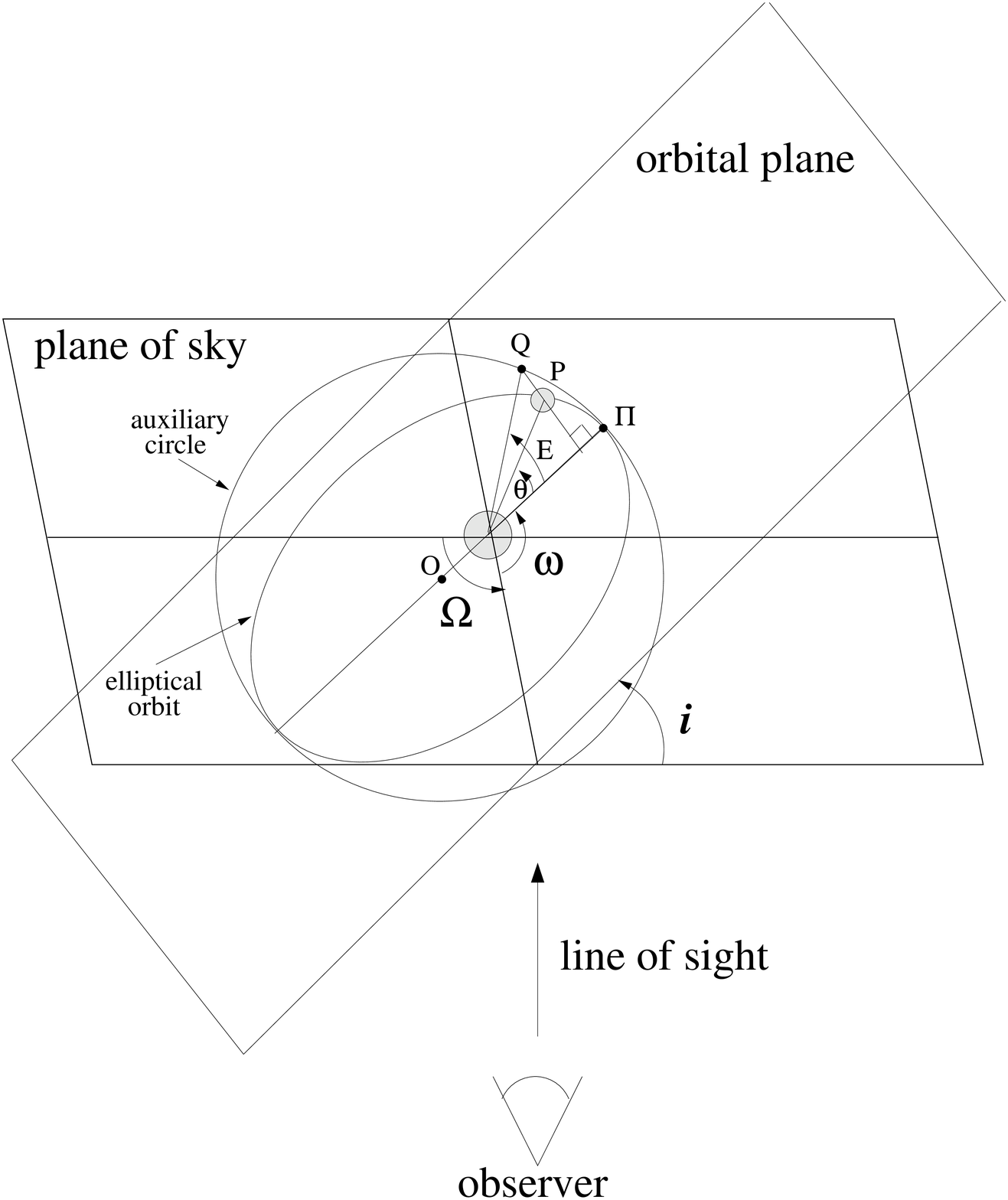}
\caption{Angles orientating the orbital plane with respect to the plane of sky,
and angles associated with the orbits. Adapted from \cite{Hilditch}.}
\label{fig:angles}
\end{figure*}

The physical model that generates light curves from eclipsing binary systems is
described by the following set of parameters: mass $M_1\in [0.5, 100]$ (in solar masses)
of the primary star (star with highest mass of the pair), mass ratio $q \in [0,1]$ (hence mass of secondary star is $M_2=qM_1$),
eccentricity $e\in [0,1]$ of the orbit and period $\rho \in [0.5,100]$ measured in days, all of which specify
the shape of the orbit. Furthermore, two angles describing the orientation of the
system are necessary \cite{Hilditch} which are known as the inclination $\imath \in [0,\frac{\pi}{2}]$
and the argument of periastron $\omega \in [0,2\pi]$ (see \figurename \ref{fig:angles}).
Inclination describes the angle between the plane of the sky and the orbital plane
and periastron is the angle $\omega \in [0,2\pi]$ that orients the major
axis of the elliptic orbit within its plane, that is $\omega$ is measured within the orbital plane. 
Finally, a third angle known as the longitude of ascending node 
($\Omega \in [0,2\pi]$)
is necessary for the complete description of a binary system. However, since it has
no effect on the observed light curves, we omit it from the model. We collectively
denote these parameters by vector $\bd{\theta}$.

The mass $M$ of each star relates to the luminosity $L$ radiated by a surface element \cite{Karttunen} of the star according to  
$L = M^{3.5}$ . Moreover, masses relate to the radii $R$ of the stars via:
\begin{equation}
R = \left\{ \begin{array}{ll}
          10^{0.053 + 0.977 \log_{10}(M)}, & \mbox{if $M < 1.728$;}\\
          10^{0.153 + 0.556 \log_{10}(M)}, & \mbox{otherwise}.\end{array} \right.
\end{equation}
These relations show that the primary star is the most luminous
one and the one with the greatest area of the pair (a star appears as a disc to an observer).
Thus, the {\em observed} area of a star is $A=\pi R^2$ and the {\em observed} luminosity is $L \pi R^2$.
Henceforth, we index quantities related to the primary star by $1$ (e.g. primary mass is $M_1$) and $2$ for the secondary star.

It is shown from Newton's laws that the orbits of an object in the gravitational field
of another object is a conic section of eccentricity $e$. 
Here we are interested in the case where $0\leq e<1$ that corresponds to closed orbits.  
We formulate two-body systems as systems where one body is fixed and the other is in orbital motion\footnote{It is
shown in \cite{Karttunen} that in the relative motion system, the eccentricity, period and 
semi-major of the moving body's orbit are equal to their counterparts in the two-body system,
and only the masses transform.}.

The position of the orbiting body is calculated
by Kepler's equation as the distance $r$ from the fixed companion star on the elliptical orbit
\cite{Hilditch},
\begin{equation}
r(t) = \frac{a(1-e^2)}{1+e\cos \theta(t)},
\label{eq:radius_orbit}
\end{equation}
where $t$ is time and $a$ is the semi-major axis of the ellipse calculated by Kepler's third law.
Point $\Pi$ in \figurename \ref{fig:angles} is the periastron, the point where
the distance between the orbiting and fixed body is minimal.
Angle $\theta$ is the angle between the radius and the periastron.
Knowledge of $\theta$ would allow us to determine
the position of the orbiting body. Angle $\theta$ is indirectly inferred via an auxiliary circle 
centered at the center of the ellipse $O$ and radius equal to semi-major axis. Point $Q$ 
is the vertical projection of the orbiting body's position $P$ to the auxiliary circle. Angle $E$
is called the eccentric anomaly and is given by Kepler's equation \cite{Hilditch}:
\begin{equation}
E(t)  =  e \sin E(t) + \frac{2\pi}{\rho}(t-\tau),
\end{equation}
%
%
%
%
where $\tau$ is the instance of time that the body was at
the periastron. Kepler's equation does not admit
an analytical solution but can be approximated through the Newton-Raphson method. 
%
%
%
%
By geometrical arguments it is shown that the relation between the true and eccentric
anomaly reads:
\begin{equation}
\tan \frac{\theta(t)}{2} = [(1+e)/(1-e)]^{\frac{1}{2}} \tan (\frac{E(t)}{2})
\label{eq:true_eccentric_relation}
\end{equation}
By knowledge of $\theta$ we can determine the position of the second star on the orbit using
(\ref{eq:radius_orbit}) and (\ref{eq:true_eccentric_relation}).
These positions correspond to the orbital plane and must be projected to the plane
of the observer in the form of Cartesian coordinates \cite{Hilditch}:
\begin{eqnarray}
X(t) &=& r(t) (\cos(\Omega) \cos(\omega+\theta(t)) - \sin(\Omega) \sin(\omega+\theta(t))  \cos(\imath)), \\
Y(t) &=& r(t) (\cos(\Omega) \cos(\omega+\theta(t)) + \cos(\Omega) \sin(\omega+\theta(t))  \cos(\imath)), \\
Z(t) &=& r(t)  \sin(\omega+\theta(t))  \sin(\imath), 
\label{eq:projections}
\end{eqnarray}
which concludes the 
determination
\footnote{The angle $\Omega$ does influence the position of the orbiting body. However, it does not have an influence on the light curve and thus we treat it as a constant $\Omega=0$.}
of positions of the stars with respect to the observer.

\begin{figure*}[!t]
\centering
\includegraphics[width=11.5cm]{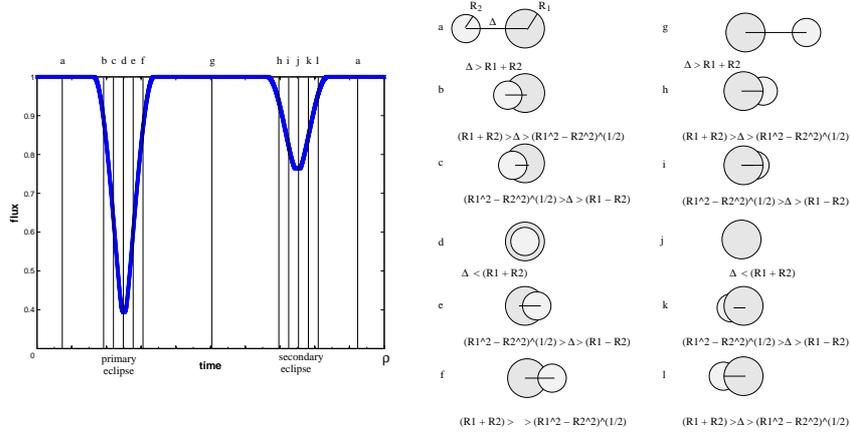}
\caption{Positions of stars (relative to observer's line of sight) and corresponding light curve phases.}
\label{fig:light_and_orbit}
\end{figure*}

An observer of the binary system receives a variable luminosity from the eclipsing binary system that plotted 
against time forms a light curve.  This variability is due to the eclipses that occur when one body passes in 
front (in the line of sight of the observer) of the other. This is illustrated in \figurename 
\ref{fig:light_and_orbit}. When no eclipse occurs (positions $a,g$) the luminosity is equal to the sum of
the luminosities radiated from the two bodies. The curved parts of the light curve occur due to partial occlusions.
Two eclipses take place at each period, one
primary eclipse (position $d$), when the most luminous body of the pair is obscured the most, and a secondary
eclipse (position $j$), when the most luminous body obscures its companion the most.

Obscured parts of the disks of the stars can be calculated via geometrical arguments.
\footnote{see http://www.physics.sfasu.edu/astro/ebstar/ebstar.html. Last access on 12-0-07.}
The obscured area of each star at time $t$ is denoted by $\Delta A_1(t)$ and $\Delta A_2(t)$.
The luminosity $f_{\bd{\theta}}(t)$ received by the observer at time $t$ depends on the luminosities $L_i$, areas $A_i$ and obscured areas\footnote{Recall that $i=1$ and $i=2$ index the primary and secondary stars, respectively} $\Delta A_i$ via
\begin{equation}
f_{\bd{\theta}}(t) = L_1 (A_1 - \Delta A_1(t))  + L_2  (A_2 - \Delta A_2(t)).
\end{equation}


\section{Noise Model for Light Curves}
\label{sec:noise_model}

Based on the physical model a probabilistic generative noise model arises naturally.
Observed light curves, denoted by $\bd{O}$, are noisy signals:
\begin{equation}
\bd{O}(t) = f_{\bd{\theta}}(t) + \epsilon(t),
\end{equation}
where $\epsilon$ is i.i.d. Gaussian noise with
variance $\sigma^2$. Thus, we  regard a light curve $\bd{O}$ of period $\rho(\mathbf{O})$ sampled at times $t \in \mathcal{T} = \{t_1=0, t_2, ..., t_T= \rho(\mathbf{O})\}$ 
as a realisation drawn from a multivariate spherical normal distribution.
We denote the noise model associated with parameters $\bd{\theta}$ 
by $p(\bd{O}|f(.;\bd{\theta}),\sigma^2)$ or simply by $p(\bd{O}|\bd{\theta})$.

\section{Model for Topographic Organisation}
\label{sec:topographic}

The starting point of our model formulation is the form of a mixture model composed
of $C$ noise models as described in section \ref{sec:noise_model}:
\begin{eqnarray}
p(\bd{O}|\bd{\Theta}) &=& \sum_{c=1}^C P(c)\ p(\bd{O} | \bd{\theta}_c),
\label{eq:GTM_Mix}
\end{eqnarray}
where $P(c)$ are the mixing coefficients, $\bd{\Theta}$ encapsulates all parameter
vectors $\lbrace \bd{\theta}_c \rbrace_{c=1:C}$ and $p(\bd{O}|\bd{\theta}_c)$
corresponds to the $c-$th model component with parameter vector $\bd{\theta}_c$.
We simplify notation $p(\bd{O} | \bd{\theta}_c)$ to $p(\bd{O} | c)$.
Assuming that dataset $\mathcal{D}$ contains $N$ independently generated fluxes  $\bd{O}\Nth{n}$,
the posterior of the $\bd{\Theta}$ is expressed as:
\begin{eqnarray}
p(\bd{\Theta}|\mathcal{D}) \propto p(\bd{\Theta}) \prod_{n=1}^N p(\bd{O}\Nth{n}|\bd{\Theta})
                              =    p(\bd{\Theta}) \prod_{n=1}^N\sum_{c=1}^C P(c) p(\bd{O}\Nth{n}|c) 
\label{eq:map_estimation}
\end{eqnarray}
where the mixing coefficients can be ignored as $P(c)=\frac{1}{C}$.

Topographic organisation is introduced in the spirit of the GTM \cite{Bishop1998} by
requiring that the component parameter vectors $\bd{\theta}_c$ correspond to a regular grid of points
$\bd{x}_c, c=1,\dots,C$, in the two dimensional latent space $\mathcal{V}=\lbrack -1,1 \rbrack^2$.
A smooth nonlinear function $\Gamma$ maps each point $\latent{}\in \mathcal{V}$
to a point $\Gamma(\latent{})$ that addresses a model $p(\cdot|\latent{})$.
Points $\Gamma(\latent{})$ are constrained on a two-dimensional manifold $\mathcal M$ that
is embedded in space $\mathcal{H}$, the space of parametrisations of our noise models.
Since the neighbourhood of $\Gamma$-images of $\bd{x}$ is preserved due to continuity of $\Gamma$,
a topographic organisation emerges for the models $p(\cdot|\latent{})$.
Function $\Gamma$ is realised as a RBF network \cite{Bishop1998}:
\begin{equation}
\Gamma(\latent{}) = \bd{W}\bd{\phi}(\latent{}),
\end{equation}
where matrix $\bd{W} \in \mathit{R}^{6 \times K}$ contains the free parameters of the model
($6$ is the number of parameters in $\{M_1,q,e,\imath,\omega,\rho\}$), and
$\bd{\phi}(.)=(\phi_1(.),...,\phi_K(.))^T, \phi_k(.):\ \mathit{R}^2\rightarrow \mathit{R}$
is an ordered set of K  nonlinear smooth basis functions. However,
this mapping may produce invalid parameter vectors, since the output of the RBF network is unbounded.
We therefore redefine mapping $\Gamma$ as:
\begin{equation}
\Gamma(\latent{}) = \bd{A}g(\bd{W}\bd{\phi}(\latent{}))+\bd{v},
\label{eq:mapping}
\end{equation}
where:

\begin{itemize}
\item  $g$ a vector-valued version of the sigmoid function that ``squashes'' each element in $\lbrack 0,1\rbrack$:
\begin{equation}
g(\bd{y}) = \bigg[ \frac{1}{1+\exp(-y_1)},\ \frac{1}{1+\exp(-y_2)},\ \dots, \frac{1}{1+\exp(-y_Y)} \bigg]^T \\,
\end{equation}
\item $\bd{A}$ is  a diagonal matrix that scales parameters to the appropriate range. $\bd{A}$ has as diagonal elements the length of range $(\theta^{max}_i-\theta^{min}_i)$ for each parameter, so that $A=diag( (100-0.5),\  (1-0),\ (1-0),\  (2\pi-0),\  (\frac{\pi}{2}-0),\  (100-0.5) )$.
\item vector $\bd{v}$ shifts the parameters to the appropriate interval. $\bd{v}$ contains the minimum value $\theta^{min}_i$ for each parameter $\theta_i$:
$\bd{v} =   \lbrack  0.5,     \   0,\   0,\     0,\    0,\       0.5 \rbrack^T$.
\end{itemize}

The redefined mapping $\Gamma$ now takes a point $\latent{}$ in space $\mathcal{V}$ to a valid 
parameter vector $\Gamma(\latent{})$ that addresses a noise model in $\mathcal{M}$.
Thus, $\bd{\Theta}$ has become a function of the weight matrix 
$\bd{W}$ of the RBF network, $\bd{\Theta}(\bd{W})$. Hence. the logarithm of the posterior from (\ref{eq:map_estimation})
now reads:
\begin{equation}
\log p(\bd{\Theta}(\bd{W})|\mathcal{D}) \propto \log p(\bd{\Theta}(\bd{W})) + \sum_{n=1}^N \log \sum_{c=1}^C p(\bd{O}\Nth{n}|\latent{c}).
\label{eq:map_estimation_GTM}
\end{equation}

\begin{figure}
\centering
\includegraphics[ width=8cm ]{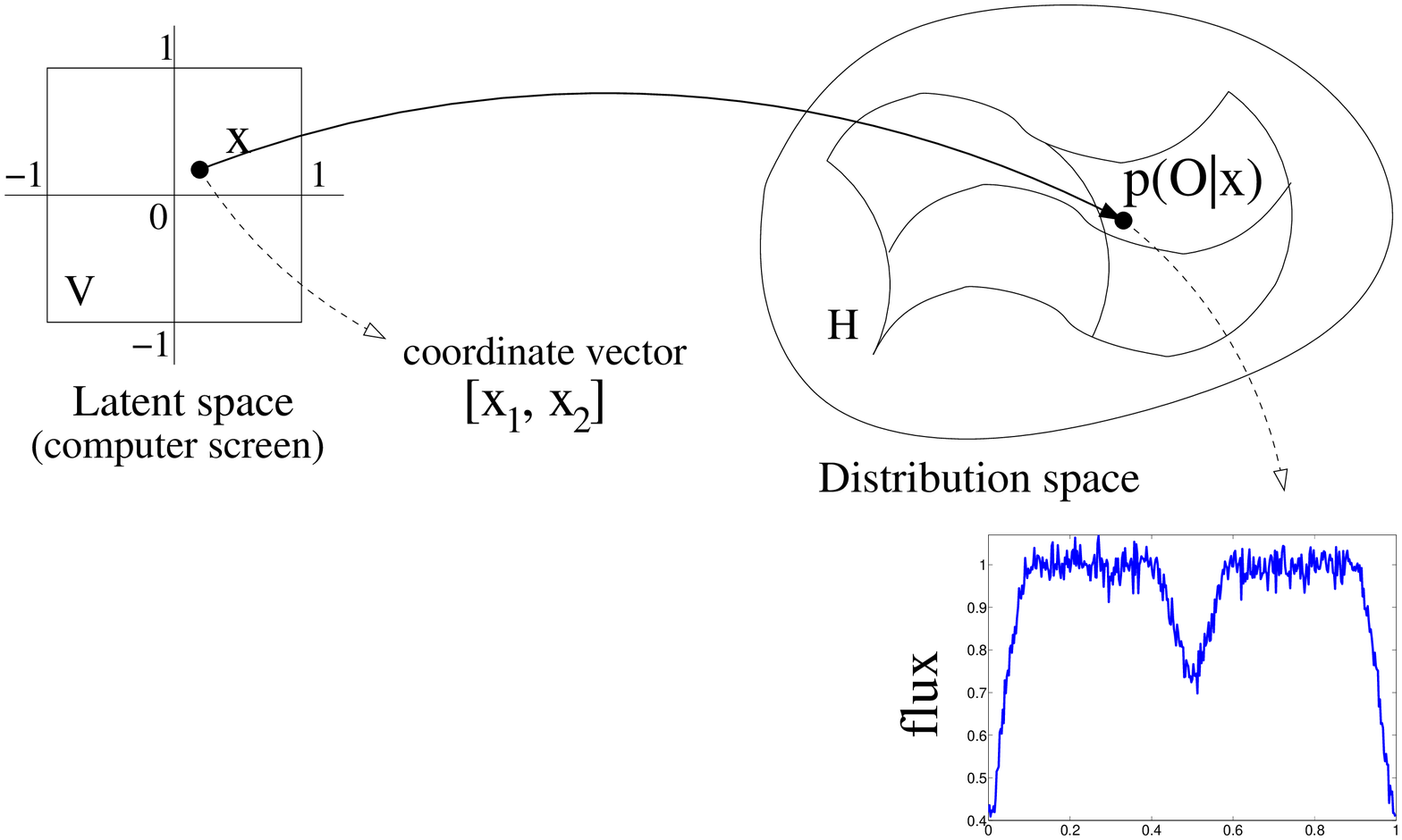}
\caption{Formulation of the topographic mapping model.}
\label{fig:model_formulation}
\end{figure}

Figure \ref{fig:model_formulation} summarises the model formulation.
Each point $\latent{}$ of the visualisation space $\mathcal{V}$ is non-linearly and smoothly mapped
via $\Gamma$ to model parameters that identify the corresponding noise model $p(\cdot | \latent{})$.
These parameters are constrained on a two-dimensional manifold $\mathcal M$ embedded in $\mathcal H$, the space of all possible parametrisations of our noise model.
In the spirit of \cite{Bishop1998}, the model can be used to visualise  observed fluxes $\bd{O}$ by calculating the posterior probability of each grid point $\bd{x}_c \in \mathcal{V}$, given  $\bd{O}$:
\begin{equation}
p(\bd{x}_c | \bd{O})
 = \frac{P(\bd{x}_c)p(\bd{O}|\bd{x}_c)} {p(\bd{O})}
 = \frac{P(\bd{x}_c)p(\bd{O}|\bd{x}_c)}
         {\sum_{c'=1}^C P(\bd{x}_{c'})p(\bd{O}|\bd{x}_{c'})} 
 = \frac{p(\bd{O}|\bd{x}_c)} {\sum_{c'=1}^C p(\bd{O}|\bd{x}_{c'})}. 
\end{equation}

Each observed flux $\bd{O}$ is then represented in the visualisation space $\mathcal{V}$ by a point $proj(\bd{O})\in \mathcal{V}$
given by the expectation of the posterior distribution over the grid points:

\begin{equation}
proj(\bd{O}) = \sum_{c=1}^C p(\latent{c}|\bd{O})\latent{c}.
\end{equation}

We train our model in the MAP estimation framework with a physically motivated prior $p(\bd{\Theta})$ obtained from relevant literature \cite{Devor2005,Halbwachs2003,Miller1979,Paczynski2006}. To that purpose we employ the EM algorithm.
Note that, due to the nature of the physical model formulation in sections \ref{sec:physical_model} and \ref{sec:noise_model},
the M-step cannot be carried out analytically, nor can the derivatives of  
expected complete-data log-posterior with respect to the RBF network parameters $\bd{W}$ be analytically obtained.
However, the EM algorithm does not necessarily require that an optimum is achieved in the M-step;  
it is sufficient that the likelihood is merely improved \cite{McLachlan2004}. 
For our purposes we resort to numerical optimisation by employing a $(1+1)$ evolutionary strategy described in \cite{Rowe2004}.
The fitness function for the evolutionary strategy is the expected complete-data log-posterior.

\section{Experiments}
\label{sec:experiments}

\subsection{Datasets}
\label{sec:datasets}

We performed experiments on two datasets. Dataset $\mathit{1}$ is a
synthetic dataset that consists of $200$ light curves (fluxes).  A common
set of model parameters, $\{ M_1=5,q=0.8,e=0.3,\imath=\frac{\pi}{2}
\}$ was defined.  However, two distinct values $\rho_1=2, \rho_2=5$ of
period and $\omega_1=0, \omega_2=\frac{5}{6}\pi$ of argument of
periastron were used, to create $4$ classes of light curves ($50$ in
each class) by the combinations of these values,
$\{\rho_1,\rho_2\}\times \{\omega_1,\omega_2\}$. The discerning
characteristic of each class is the position of each secondary eclipse
and the widths of the eclipses. Each light curve was then generated
from these four ``prototypical'' parameter settings corrupted by a Gaussian noise.
Gaussian noise was also subsequently added to the generated light curves to simulate observational errors.

Dataset $\mathit{2}$ consists of light curves from real observations
obtained from two resources available\footnote{Last accessed on the
12th September 2007.} on the WWW: the {\it Catalogue and Archive of
Eclipsing Binaries} at http://ebola.eastern.edu/  and the {\it All Sky
Automated Survey}. Dataset $\mathit{2}$ was preprocessed before training using local linear interpolations.
Preprocessing is necessary as one needs to account for gaps in the monitoring process
and for overlapping observations.
Light curves must also be phase-shifted so that their first
point is the primary eclipse and resampled to equal length as described in section \ref{sec:noise_model}.
Finally, the light curves were resampled at $T=100$ regular intervals which
was judged an adequate sample rate.

\subsection{Training}
\label{sec:training}

The lattice was a $10 \times 10$ regular grid (i.e. $C=100$) and the RBF network consisted of $M=17$ basis
functions; 16 of them were Gaussian radial basis functions of variance $\sigma^2=1$ centred on a $4 \times 4$
regular grid in $\mathcal{V}=[0,1]^2$, and one was a bias term.
The variance of the observation noise in the local models $p(\bd{O} | \latent{})$ was set to $\sigma^2=0.075$.

\subsection{Results}
\label{sec:results}

\begin{figure}
\centering
\fbox{\includegraphics[width=5.5cm ]{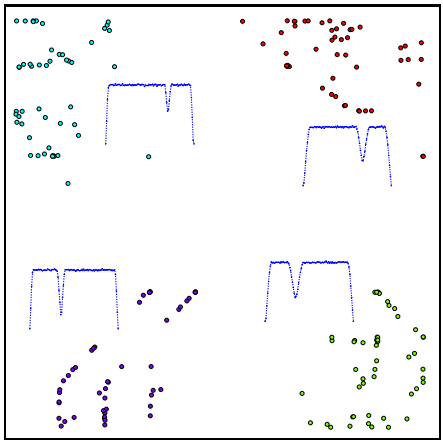}}
\caption{Visualisation of synthetic dataset. A representative light curve is plotted next to each cluster.
}
\label{fig:map_toy_fluxes}
\end{figure}

\figurename \ref{fig:map_toy_fluxes} presents the topographic map constructed
for the synthetic dataset. Each point stands for a light curve projected to latent visualisation space $\mathcal V$
and is coloured according to class membership. The class memberships of synthetic fluxes were not used during the training process.
Also, next to each cluster, a typical light curve has been plotted.
The classes have been identified and organised appropriately, each 
occupying one of the four corners of the plot.

\begin{figure}
\centering
\fbox{\includegraphics[width=7.5cm]{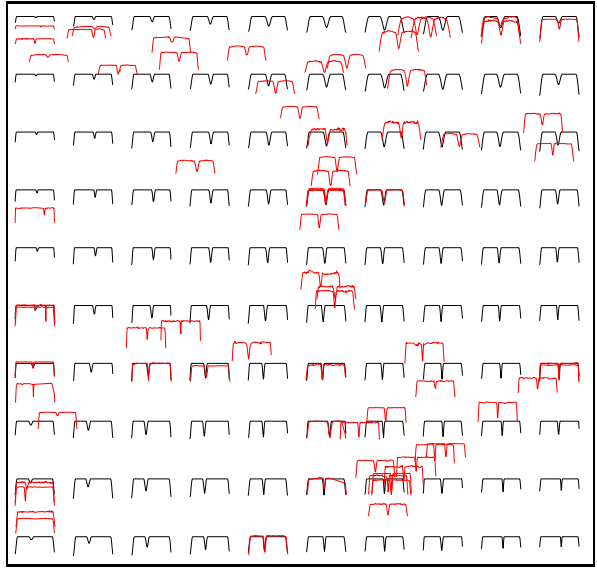}}
\caption{Visualisation of dataset $\mathit{2}$ of real data. Light curves in red are the projected real data
and light curves in black are the light curves of the underlying local noise models.}
\label{fig:map_real_fluxes}
\end{figure}

\figurename \ref{fig:map_real_fluxes} presents the topographic map constructed
for the dataset  of real observed light curves. The red curves are the data projected against
the underlying local noise models displayed in black.
Several interesting observations can be made about the topographic formation of the light curves on the resulting map.
In the lower right-hand corner binary systems of large periods are found. 
The median period of the systems in our sample is 2.7~days, and binaries like {\it V459 Cas}, with a
period of 8.45~days lie in this corner. 
Systems with short period have the appearance
of a wide V-shaped eclipse in the shape of their light curve, and
inhabit the top and left edges of the map, e.g.  WY~Hya (Period: 0.7 days) and RT~And (Period: 0.6 days).
At the lower left of the map, we find systems with high eccentricity, e.g. V1647~Sgr. High eccentricity causes the light curve to appear
assymetric, so that the period of the eclipse occurs further and further away from the center. 
On the other hand, very symmetric curves indicate orbits of low eccentricity (more circular)
and low mass-ratio (stars of similar mass), and indeed we find systems like DM~Vir ($e=0.03$, mass ratio=1) 
and CD~Tau  ($e=0.0$, mass ratio=1.05) in the cluster in the lower-right hand corner of the map. 
Finally, low-inclination systems, occupy the top left-hand corner of the map, and these orbits will have very shallow eclipses
as the companion star barely eclipses the primary star. 

\section{Conclusions}

We have presented a model-based probabilistic approach
for the visualisation of eclipsing binary systems. The model is formulated as a constrained-mixture
of physically motivated noise models. 
As a consequence, a clear cost function naturally arises which drives the optimisation of the model.
In our experiments we have demonstrated that the resulting maps can be interpreted in a transparent way
by inspecting the underlying local noise models. Furthermore, modification and refinement of the local noise models is possible,
to account for greater physical fidelity by incorporating physical
aspects for non-spherical stars and even more sophisticated phenomena
such as gravity darkening.

\bibliographystyle{splncs}
\bibliography{icann09}
\end{document}

%% file: CommandDefs.tex
\renewcommand{\figurename}{Fig.\ }

\newcommand{\bd}[1]{\mbox{\boldmath $#1$}}
\newcommand{\RR}[1] {\mathit{R}^{#1}} 
\newcommand{\kronDelta}[2]{\delta_{#1,#2}}
\newcommand{\dx}[2] {\frac{\partial #2} {\partial x_{#1}}}
\newcommand{\dxdx}[3] {\frac{\partial^{2}#3} {\partial x_{#1} x_{#2}}}
\newcommand{\Nth}[1]{^{(#1)}}
\newcommand{\nth}[1]{_{(#1)}}
\newcommand{\vect} {\bd{t}}
\newcommand{\vectN}[1] {\bd{t}\Nth{#1}}
\newcommand{\GaussianPDF}[3] { {\mathcal N}(#1;#2,#3)}
\newcommand{\KLD}[2] {D_{KL}\lbrack #1 \vert\vert #2 \rbrack}
\newcommand{\argmax}[2] {\mathop{argmax}_{#1}\bigg\{#2\bigg\}}
\newcommand{\argmin}[2] {\mathop{argmin}_{#1}\bigg\{#2\bigg\}}
\newcommand{\expect}[1] {E\lbrack {#1} \rbrack}
\newcommand{\expectWRT}[2] {E_{#1} \lbrack {#2} \rbrack}

\newcommand{\latent}[1]{\bd{x}_{#1}}
\newcommand{\model}[1]{\mbox{$p(\cdot|#1)$}}

\newcommand{\Seq} {\bd{S}}
\newcommand{\seq} {\bd{s}}
\newcommand{\SeqN}[1] {\bd{S}\Nth{#1}}
\newcommand{\seqN}[1] {\bd{s}\Nth{#1}}

\newcommand{\Symb}[1] {\bd{S}_{#1}}
\newcommand{\symb}[1] {\bd{s}_{#1}}
\newcommand{\SymbN}[2] {\bd{S}_{#1}\Nth{#2}}
\newcommand{\symbN}[2] {\bd{s}_{#1}\Nth{#2}}

\newcommand{\Zcomp}[2] {z^{(#1)}_{#2}}
\newcommand{\Zstate}[3] {z^{(#1,#2)}_{#3}}
\newcommand{\Ztrans}[4] {z^{(#1,#2)}_{#3 \rightarrow #4}}
\newcommand{\State}[1]{Q_{#1}}
\newcommand{\state}[1]{q_{#1}}

\newcommand{\Tree}{\bd{Y}}
\newcommand{\tree}{\bd{y}}
\newcommand{\TreeN}[1]{\bd{Y}\Nth{#1}}
\newcommand{\treeN}[1]{\bd{y}\Nth{#1}}

\newcommand{\Subtree}[1]{\bd{Y}_{#1}}
\newcommand{\subtree}[1]{\bd{y}_{#1}}
\newcommand{\SubtreeN}[2]{\bd{Y}_{#1}\Nth{#2}}
\newcommand{\subtreeN}[2]{\bd{y}_{#1}\Nth{#2}}

\newcommand{\parent}[1]{\rho{(#1)}}
\newcommand{\child}[1]{c{(#1)}}

\newcommand{\NodeLabel}[1]{\bd{O}_{#1}}
\newcommand{\nodeLabel}[1]{\bd{o}_{#1}}
\newcommand{\NodeLabelN}[2]{\bd{O}_{#1}\Nth{#2}}
\newcommand{\nodeLabelN}[2]{\bd{o}_{#1}\Nth{#2}}

\newcommand{\SubTreeExcl}[2]{\bd{Y}_{#1 \backslash #2}}
\newcommand{\subTreeExcl}[2]{\bd{y}_{#1 \backslash #2}}
\newcommand{\SubTreeExclN}[3]{\bd{Y}_{#1 \backslash #2}\Nth{#3}}
\newcommand{\subTreeExclN}[3]{\bd{y}_{#1 \backslash #2}\Nth{#3}}

\newcommand{\Ainit}{\bd{A}^{(\bd{\pi})}}
\newcommand{\Atrans}[1]{\bd{A}^{(\bd{T}_{#1})}}
\newcommand{\Aemiss}[1]{\bd{A}^{(\bd{B}_{#1})}}